# Implement and commissioning of the beam energy feedback system in BEPCII linac

Shaozhe Wang(王少哲)[1;2;1)] Yunlong Chi(池云龙)[2] Rong Liu (刘熔)[2] Xuefang Huang (黄雪芳)[1;2]
Lei Qian (钱磊)[2]

1 University of Chinese Academy of Sciences, Beijing 100049, China
2 Institute of High Energy Physics, Chinese Academy of Sciences, Beijing 100049, China

**Abstract**：In order to ensure the beam quality and meet the requirements introduced by the BEPCII storage ring, the beam energy feedback system has been developed at the exit of the linac. This paper describes the implementation and commissioning of this system in detail. The energy feedback system consists of an energy measurement unit, an application software and an execution unit. In order to ensure the real-time monitoring and adjustment of beam energy, we need to introduce a non-interceptive type of online beam energy measurement method which is on the first try in China and the effective mechanism of energy adjustment to achieve this goal. The adjustment of energy is achieved by adjusting the output microwave phase of the RF power source station. The system was put into operation in March 16th, 2016 and achieved the desired results. It can effectively eliminate the low point of the injection rate caused by the fluctuation of the beam center energy and has played an important role in maintaining a high constant injection rate.

**Key words:** linac, energy feedback, electron, positron, BEPCII, BPM, phasing system

**PACS:** 29.20.Ej

## 1. Introduction

After Beijing electron positron collider (BEPC) being upgraded into Beijing electron positron collider II (BEPCII), its storage ring has higher demand on the quality of the beam at the exit of the linac. The main parameters of the injection beam of BEPC and BEPCII have been listed in Table1 [1]. The beam energy dispersion consists of two parts, one is the dispersion caused by the fluctuation of the accelerating phase, the other is the dispersion caused by the change of the beam center energy. However, for the long-term stability of the beam, the latter has greater influence.

Table 1. Main parameters of the injection beam of BEPCII.

| Parameter | | BEPC | BEPCII | Unit |
|---|---|---|---|---|
| Energy | | 1.3 | 1.89 | GeV |
| Pulse width | | 2.5 | 1.0 | ns |
| Bunch length ($2\sigma_z$) | | 10 | 10 | ps |
| Beam current | ($e^+$) | 4.5 | 37 | mA |
| | ($e^-$) | 1 | 1 | A |
| Repetition frequency | | 12.5 | 50 | Hz |
| Beam emittance | ($e^+$) | 1.7 | 1.6 | mm·mrad |
| | ($e^-$) | -- | 0.2 | mm·mrad |
| Energy dispersion | ($e^+$) | ±0.8 | ±0.5 | % |
| | ($e^-$) | ±0.5 | ±0.5 | % |

In the long time operation of the linac, the drift of the center energy cannot be avoided. When the beam energy is out of the range, the injection rate will drop rapidly. To suppress the fluctuation of the beam center energy at the exit of the linac and make the whole accelerator run stably, a beam energy feedback system has been developed. It also passed the online testing on March 16th, 2016, and from then on, came into use. This system compensates the drift of the beam center energy by making a local adjustment of the energy gain. The klystrons in BEPCII linac are all working on the saturation condition, their changes of output power resulted from the variation of driving voltage are very small [2]. Hence, the only thing we should do to change the energy gain is to adjust the accelerating phase of klystron.

To realize the beam energy feedback system, we first set up a kind of non-interceptive online beam energy measurement mechanism at the exit of linac by using Experimental Physics and Industrial Control System (EPICS) based Input/Output controller (IOC). Then the measurement results together with the target values are sent to a Graphical User Interface (GUI) application. In the GUI application software, the controlled quantity is calculated and sent to the execution unit in phasing system. We choose the accelerating phase of RF power source station No.16, the last klystron at the end of linac, as the controlled object. And it can be estimated that the increase of the energy dispersion because of the adjustment is less than 0.015% at the inject energy of 1.89 GeV. Under the help of the phasing system, the energy adjustment can be achieved. The structure diagram of the beam energy feedback system is shown in Fig. 1. This paper presents the implement of the beam energy feedback system for BEPCII linear accelerator and its performance.

---

1) E-mail: wangsz@ihep.ac.cn

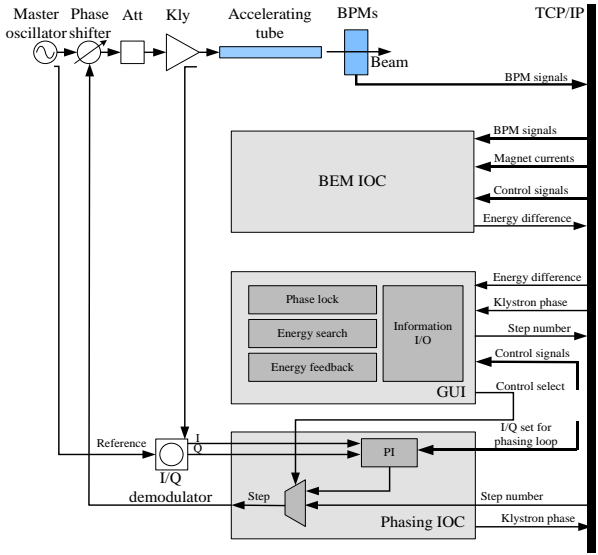

Fig. 1. The structure diagram of the beam energy feedback system in BEPCII linac.

## 2. Beam energy measurement mechanism

Different from other measurement methods [3–5], the beam energy measurement mechanism in BEPCII linac is an online measurement which eliminates the effects of beam orbit fluctuation for the accuracy and does not affect the beam for its injection task. It uses three stripline beam position monitors (BPMs) along with the beam orbit to measure the difference between the real beam energy and the nominal energy. The first two BPMs are in the end of the straight common transmission line and their task is to detect the beam orbit. The third one is in the big dispersion place of the transmission line following the bending magnets. The layout of the BPMs at the end of BEPCII linac is shown in Fig. 2. There are two groups of BPMs, TE/PBPM1 and TE/PBPM3, in the big dispersion place that can be used to do the energy measurement. A correlation test has been done for making sure which group is more reliable to do the measurement. During the test, it shows that the displacement of the beam center measured by TE/PBPM3 became smaller instead of bigger as the beam orbit has a larger offset. Along with the phenomenon, there is an abnormal decrease of the ADC converting signals [6]. What's more, between the second BPM and TE/PBPM3 there are more elements which can lead to bigger error accumulation in computing. Hence, TE/PBPM1 is selected as the third BPM for energy measurement.

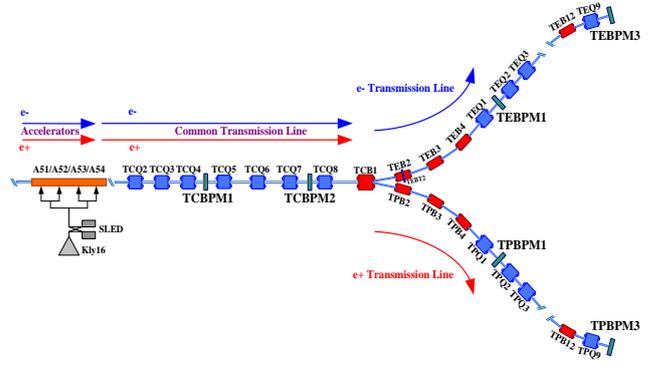

Fig. 2. The layout of the BPMs at the end of BEPCII linac.

If the electron beam goes through TCBPM1, TCBPM2 and TEBPM1 in sequence and their horizontal positions are recorded as $x_1$, $x_2$ and $x_3$, respectively by the BPMs, the relative energy bias can be written as [7]

$$\frac{\Delta E}{E} = k_1 x_1 + k_2 x_2 + k_3 x_3, \quad (1)$$

where $k_1$, $k_2$, $k_3$ are the coefficients depending on the transfer matrices between the BPMs. Eq. (1) considers the effect of orbit fluctuation, and is much weaker in terms of the limited conditions compared with the classical equation:

$$\frac{\Delta E}{E} = \frac{\Delta x}{\eta}, \quad (2)$$

where $\Delta x$ is the displacement of the beam center in the horizontal direction, $\eta$ is the value of dispersion function at that place. For instance of BEPCII, when the designed energy of electron is 1.89 GeV and the positions recorded by the BPMs are -1mm, -1mm, and 1mm, $\Delta E$ is about 0.86 MeV according to Eq. (2) with $\eta$ 2200mm. However, by using Eq. (1) which takes the orbit fluctuation into account, $\Delta E$ is about -0.68 MeV.

The error analysis of this method has also been done. Table2 [8] lists the maximal installation tolerances of the quadrupole and bending magnets and other related errors which may lead to the error accumulation during calculation. According to the data in Table2, the resolution of this method has been estimated and the result is 0.26 MeV at 2.5 GeV [7], which meets the requirement of the beam energy feedback system.

Table 2. Summary of the maximal installation tolerances of the quadrupole and bending magnets and other related errors.

| The installation tolerance of the quadrupole lens | | | |
|---|---|---|---|
| **Name** | **Value** | **Name** | **Value** |
| $|\Delta x|$ | 0.2 mm | $|\Delta\theta_x|$ | 1 mrad |
| $|\Delta y|$ | 0.2 mm | $|\Delta\theta_y|$ | 1 mrad |
| $|\Delta z|$ | 1.0 mm | $|\Delta\theta_z|$ | 2 mrad |
| **The installation tolerance of the bending magnet (from B-1 to B-17)** | | | |
| **Name** | **Value** | **Name** | **Value** |
| $|\Delta x|$ | 0.5 mm | $|\Delta\theta_x|$ | 0.5 mrad |
| $|\Delta y|$ | 0.5 mm | $|\Delta\theta_y|$ | 0.5 mrad |
| $|\Delta z|$ | 1.0 mm | $|\Delta\theta_z|$ | 0.6 mrad |
| **Other data** | | | |
| Name | | | Value |
| Stability of the quadrupole lens | | | 0.001 |
| Stability of the bending magnet | | | 3 *0.0001 |
| BPM resolution | | | 0.05 mm |
| BPM position calibration | | | 0.05 mm |
| BPM installation tolerance | | | 1 mm |
| Field gradient interpolation error | | | 0.001 |

The realization of the online measurement is in a beam energy measurement (BEM) IOC. The BEM IOC acquires the positions measured by the BPMs together with the magnet currents between them from the relevant channels of EPICS, calculates the energy difference by using certain programs, and sends the results to the Ethernet via EPICS channels.

## 3. The GUI application

The GUI application plays a significant role in this feedback system for it has two functions. One is to communicate with the users. The other is to calculate the controlled quantity and send it to EPICS channel to change the accelerating phase for energy gain adjustment.

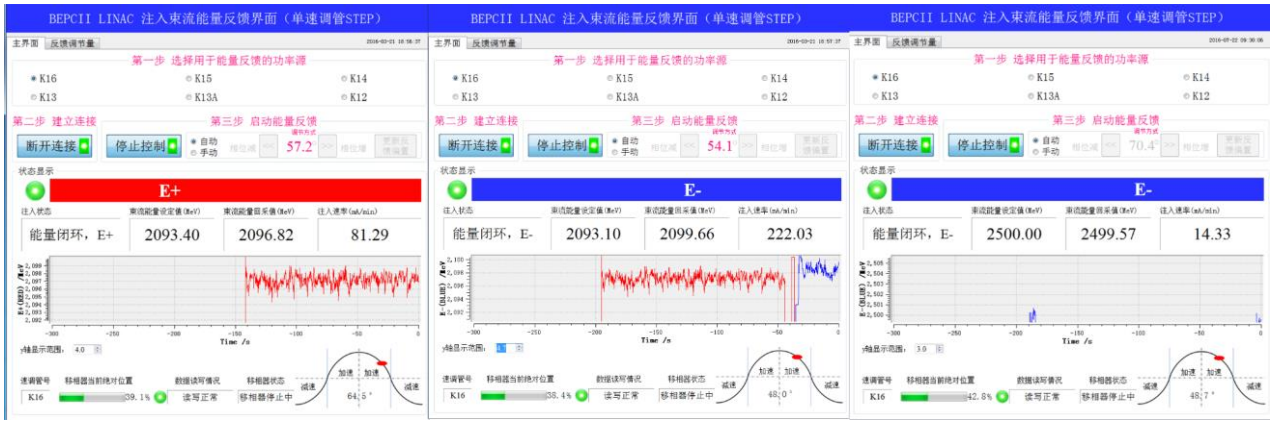

Fig. 3. The screenshot of the system in collision mode (left is for positron injection and middle is for electron injection) and in synchrotron radiation mode (right).

Developed on a platform named Qt, the application is designed to be an easy-to-use interface. Its language is Chinese for its users are all from China. The screenshot is shown in Fig. 3. On the top are the control buttons and only three steps are needed to start this system. The information of injection is shown in the middle. The accelerating phase of certain klystron is displayed by using the animation in the bottom right corner. Together with it, there is the information about the klystron and its phase shifter.

All of the algorithms, the energy searching algorithm, energy feedback algorithm and phase locking algorithm, are running in background. During the injection, the energy searching algorithm or energy feedback algorithm is activated. When the injection is finished or the phase adjustment is completed, the phase locking algorithm instead of the first two is activated to keep the phase in a certain value. The control logic of the GUI application during the injection is displayed in Fig. 4. When the ADC conversion signal of the BPM is less than 500 (the maximum is 32767), which means there is no beam going through the transmission line, the energy searching algorithm instead of energy feedback algorithm is activated. The step number is set to be 0 and the motor is driven in a constant speed of 3000 steps/s. The search range is from 0° to 90° of the accelerating phase. If the ADC conversion signal doesn't increase to 500 in the search range, the control of the application will be forbidden and a warning window will popup. Otherwise, the energy searching will be stopped and the energy feedback will start to work. If the measurement result of TEBPM1 is beyond the range of -2mm to 5.5mm or the result of TPBPM1 is beyond the range of -9mm to 7mm, the energy feedback algorithm will set the step number to be a constant value. This is because though there is beam going through the transmission line, the beam offset is still large and the BPM's noise is also too large that the result cannot be used to do the feedback. In this situation, the direction of the energy difference, positive or negative, is the only thing that can be confirmed, and the constant step number is a secure control strategy. When the ADC conversion signal is large enough and the beam offset is in the range of the BPMs, the energy feedback algorithm will set the step number under the relationship between the accelerating phase and the energy gain. Of course, it is non-linear. In the algorithm of phase locking and energy feedback, we use step number instead of speed to drive the step motor for the reason that the network between the motion control board and the PC in control room disconnects sometimes because of the disturbance of strong electromagnetic field. The step number driving mode can avoid the step motor being out of control when it cannot receive the stop signal.

## 4. The execution unit

The phase control mechanism of the phasing system in BEPCII linac [9-10] is used as an execution unit by the beam energy feedback system to adjust the accelerating phase. It consists of an isolator, a motorized phase shifter and a continuously variable attenuator. The phase shifter range is about 540° with a minimum phase step of 0.1°. The step motor of the phase shifter is controlled by a motion control card (KPCI882) and the KPCI882 card is driven by the EPICS based control software called phasing IOC.

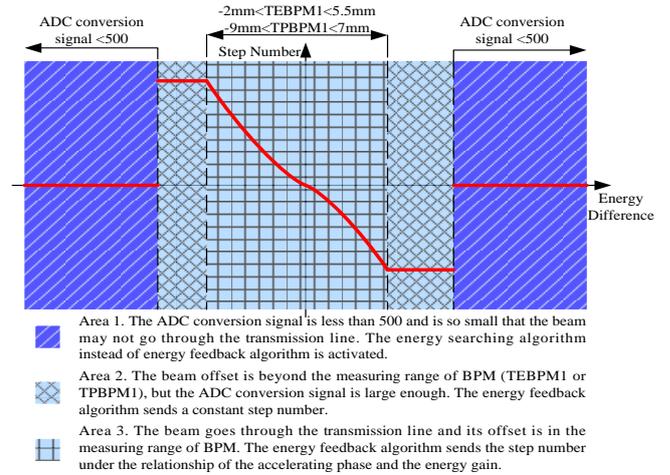

Fig. 4. The control logic of the GUI application during the injection.

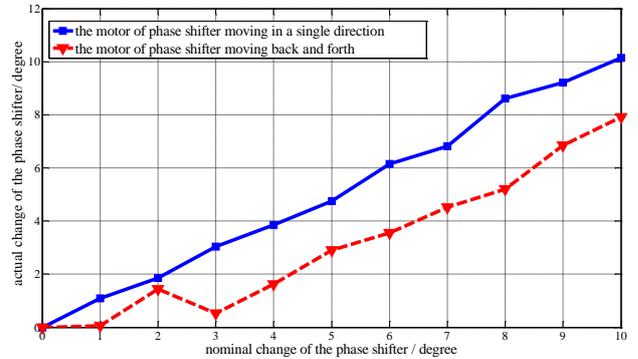

Fig. 5. The measurement result of the phase shifter return difference in klystron station No.16.

To make the phase control mechanism meet requirements of the beam energy feedback system, several improvements have been done. Firstly, the priority of beam energy feedback and phase control has been added in the phasing system. The energy feedback system has the higher priority. When it works, the phase control application of the klystron which is selected by the energy feedback system is forbidden. Secondly, the speed of the motor or in other words the impulse frequency output from KPCI882 card has been optimized. Different kinds of speed are used for different functions. For manual adjustment, the speed is set to be 1000 steps/s or nearly 2.7°/s. For energy searching, the speed is 3000 steps/s or nearly 8.0°/s, which means the searching function can be finished in less than 15 seconds. And for energy feedback, the speed is 2000 steps/s or nearly 5.3°/s. The motor speed cannot be too fast because there is pulse-width discriminating algorithm in the motor driver telling the pulses output by KPCI882 card from the interfering signal. The threshold value of the pulse-width is 100μs.

What's more, it is the first time that the return difference of the phase shifter is taken into account in phase adjustment. An EPICS based program was written to do the measurement. The result of the phase shifter in klystron station No.16 is displayed in Fig. 5. As the solid line in Fig. 5, if the motor moves in the single direction, the nominal change of the phase shifter is nearly equal to the actual change. For example, if the motor moves in the positive direction last time and is made to move 375 steps (1.0° nominally) in the same direction this time, the actual change of the phase shifter this time is about 1.0°. However, the dash line shows that if the motor moves back and forth, which means in one direction this time and in the other direction next time, there is no linear relation between the nominal change and the actual one until the nominal change is bigger than 3.0° (at this time, the actual change is 0.5°). To make the control more accurate, the direction of each time is recorded in the IOC and the step number is set according to the linear relation shown in Fig. 4. In the case of moving back and forth, if the phase shifter should be changed less than 0.5°, the step number is set to be 0 because the adjustment is meaningless in this situation. By using the new control method, the peak to peak value of the phase control decreases by nearly 50%.

## 5. Commissioning and performances of the system

An online testing for the energy feedback system has been done on March 16, 2016. The result shows that this system can effectively suppress the fluctuation of both the beam energy and the injection rate. The comparison of the injection conditions without and with the energy feedback system is displayed in Fig. 6. The left half graph is the record during the positron injection with the energy feedback system off, the right one is the record with the system on. By using the energy feedback system, the peak to peak value of the injection rate can be decreased to less than 10 mA/min. While without the system, this value is larger than 20 mA/min. In terms of beam center energy, for example, during positron injection, the target energy is set to be 2096.1 MeV and the average value is 2096.2 MeV. Meanwhile, its standard deviation is 0.8 MeV and peak to peak value is 4 MeV. The energy fluctuation is ±0.95‰. During electron injection, the target energy is 2098.2 MeV and the average value of the actual energy is 2098.2 MeV. Its standard deviation is 0.7 MeV and peak to peak value is 3 MeV. The energy fluctuation is ±0.72‰. Both the positron injection and the electron injection meet the design requirement that the energy fluctuation is about ±1‰.

This energy feedback system comes into use after passing the online testing. By adding the switching signals of electronic gun and the kickers and the ready signal of the target for positron as the trigger signals, it has adapted multiple injection modes, such as the electron injection for collision mode, the positron injection for collision mode and the electron injection for synchrotron radiation mode. Fig. 3 is the screenshot of the system in collision mode (left is for positron injection and middle is for electron injection) and in synchrotron radiation mode (right).

## 6. Summery

The beam energy feedback system for BEPCII linac has been developed successfully. It has been running stably after passing the online testing on March 16th, 2016 and has adapted multiple injection modes. The BEM IOC, the GUI application and the phasing system communicate with each other through the local area network, and work together to complete the beam center energy adjustment. This system can effectively eliminate the low point of the injection rate caused by the fluctuation of the beam center energy and has a significant contribution to increase the average rate of injection. In the future, some other functions will be added to this system to make it more powerful. For example, the IOC will monitor all the klystrons in the linear accelerator and estimate the total energy contribution in advance. In addition, the energy feedback system based on the digital LLRF system is in the research.

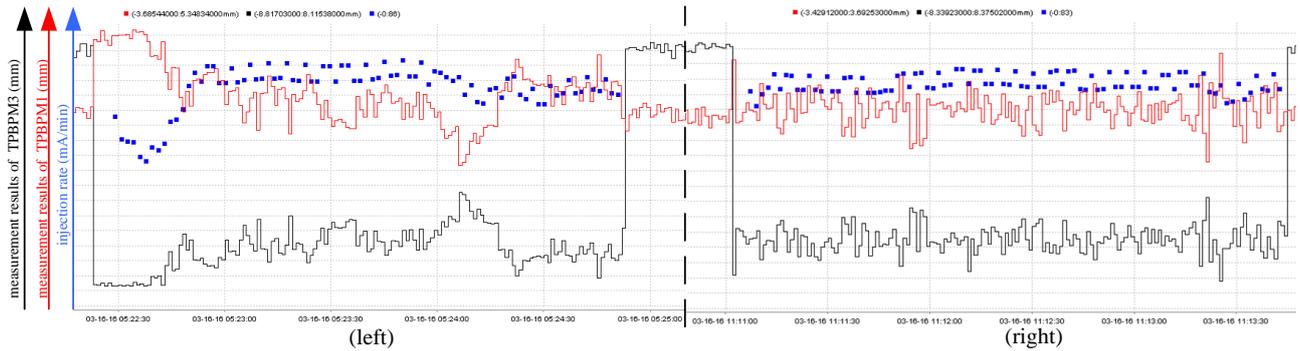

Fig. 6. The comparison of the injection without (left) and with (right) the beam energy feedback system. The blue points are the injection rate values recorded by Archive Database. The red (up) line is the measurement results of TPBPM1 and the black (down) line is the measurement results of TPBPM3. Both the red (up) line and the black (down) line indicate the fluctuation of the beam energy to some extent.